\documentclass[11pt]{article}

\usepackage[T1]{fontenc}
\usepackage[utf8]{inputenc}
\usepackage{amsmath,amssymb}
\usepackage{graphicx}
\usepackage{multicol}
\usepackage{hyperref}
\usepackage{orcidlink}
\usepackage{cite}
\usepackage{geometry}
\usepackage{float}
\title{\bf A Nonlinear Mechanism for Transient Anomalous Diffusion}

\author{
Gabriel Barreiro\,\orcidlink{0009-0003-8422-9557} \\
\small Instituto de Física, Facultad de Ciencias, Universidad Autónoma de Santo Domingo,\\
\small Avenida Alma Máter, Ciudad Universitaria, Código Postal 10105, Santo Domingo, Dominican Republic
\and
Vladimir Pérez-Veloz\,\orcidlink{0000-0003-0764-4223} \\
\small Instituto de Física, Facultad de Ciencias, Universidad Autónoma de Santo Domingo,\\
\small Avenida Alma Máter, Ciudad Universitaria, Código Postal 10105, Santo Domingo, Dominican Republic
}

\date{}

\begin{document}
\maketitle

% ==================================================
% Abstract
% ==================================================
\begin{abstract}
Diffusion is a fundamental physical phenomenon with critical applications in fields such as metallurgy, cell biology, and population dynamics. While standard diffusion is well-understood, anomalous diffusion often requires complex non-local models. This paper investigates a nonlinear diffusion equation where the diffusion coefficient is linearly dependent on concentration. We demonstrate through a perturbative analysis that this physically-grounded model exhibits transient anomalous diffusion. The system displays a clear crossover from an initial subdiffusive regime to standard Fickian behavior at long times. This result establishes an important mechanism for transient anomalous diffusion that arises purely from local interactions, providing an intuitive alternative to models based on fractional calculus or non-local memory effects.
\end{abstract}


\begin{multicols}{2}

% ==================================================
% Introduction
% ==================================================
\section{Introduction}
Diffusion is a fundamental physical process describing the net movement of particles from a region of higher concentration to one of lower concentration, driven by random thermal motion. The diffusing substance is referred to as the solute, and the medium through which it moves is the solvent.

The applications of diffusion are vast, making it a cornerstone concept in both science and engineering. For instance, understanding how carbon impurities diffuse through iron is critical for manufacturing high-quality steel \cite{porter2009phase} and understanding the movement of fluid in porous media \cite{aronson1986porous, vazquez2006porous}. In biology, the transport of nutrients within cells and the movement of ions across cell membranes are diffusion-driven processes of paramount importance~\cite{alberts2014molecular,nelson2017lehninger}. Diffusion models are also powerful tools in ecology and sociology for studying the dynamics of animal and human populations~\cite{murray2002mathematical, okubo2001diffusion}.

\subsection{Classification of Diffusion Processes}
Diffusion processes are broadly classified into two main types based on the time-dependence of their mean squared displacement (MSD) \cite{metzler2000random,hofling2013anomalous}.
\begin{itemize}
    \item Fickian Diffusion: This is the classical diffusion process governed by Fick's laws. The key characteristic is that the MSD, $\langle x^2\rangle$, is directly proportional to time, $t$:
    $$\langle x^2\rangle \propto t.$$
    \item Anomalous Diffusion: This process deviates from the classical model and is often described using the mathematical framework of fractional calculus\cite{podlubny1999fractional, hilfer2000applications}. The MSD follows a power-law relationship with time:
  $$\langle x^2\rangle \propto t^\alpha,$$
  where $\alpha$ is the anomalous diffusion exponent. When $0 < \alpha < 1$, the process is termed \emph{subdiffusion}. If $\alpha > 1$, it is \emph{superdiffusion}. The classical Fickian case is recovered when $\alpha = 1$. Anomalous diffusion is used to describe situations where there are obstacles or restriction of movement\cite{saxton1994anomalous, aronson1986porous}.
\end{itemize}
Anomalous diffusion has been observed in diverse systems ranging from in crowded biological environments \cite{weiss2004anomalous} to financial markets \cite{mantegna1995scaling} and tracer particles in plasma physics \cite{carreras2001anomalous}. However, the underlying mechanisms remain a subject of active research.

\subsection{Motivation for a Concentration-Dependent Model}
This paper investigates a model where the diffusion coefficient depends on the concentration of the solute. The motivation is straightforward: in a crowded environment, a particle's movement is likely influenced by its neighbors. Microscopic interactions—whether attractive or repulsive—between solute particles, or between solute and solvent, can lead to complex macroscopic behavior that a constant diffusion coefficient cannot capture \cite{hanggi1990reaction,redner2001guide}.

In population dynamics, for instance, migration patterns can depend on population density. Individuals might be attracted to or repelled from crowded areas, a behavior that can be modeled by a density-dependent diffusion term \cite{turchin1998quantitative}. Similarly, in cellular biology, protein diffusion in the cytoplasm is known to be concentration-dependent due to molecular crowding effects \cite{weiss2004anomalous}.

A key advantage of this approach is that it preserves the \emph{locality of interactions}. In contrast, models based on fractional calculus often encode memory effects or non-local dynamics, which can be difficult to justify from a first-principles physical standpoint \cite{west2003physics}. The model presented here provides a mechanism for anomalous behavior that avoids this conceptual difficulty, relying instead on a simple, local nonlinearity.

Recent experimental studies have indeed shown that concentration-dependent diffusion can arise naturally in various systems, from colloidal suspensions \cite{kegel2000direct} to biological membranes \cite{jacobson1995revisiting}, supporting the physical relevance of our approach.

% ==================================================
% Model Description
% ==================================================
\section{Model Description}
We begin with the one-dimensional nonlinear diffusion equation, often referred to as the Fickian diffusion equation with a concentration-dependent diffusivity:
$$\dfrac{\partial \psi(x,t)}{\partial t} = \dfrac{\partial}{\partial x}\left[D(\psi)\dfrac{\partial \psi}{\partial x}\right]$$
where $\psi(x,t)$ represents the concentration of a substance at position $x$ and time $t$, and $D(\psi)$ is the concentration-dependent diffusion coefficient.

The diffusion coefficient can be expressed as a power series in terms of the concentration $\psi$:
$$D(\psi) = \sum_{n=0}^\infty D_n\psi^n,$$
where $D_n$ are constant coefficients. If all coefficients $D_n$ for $n>0$ are zero, this expression reduces to the standard linear diffusion equation.

Substituting the series expansion into the diffusion equation yields:
$$\dfrac{\partial \psi}{\partial t} = \sum_{n=0}^\infty\dfrac{D_n}{n+1} \dfrac{\partial^2 \psi^{n+1}}{\partial x^2}.$$

In this work, we consider a diffusion coefficient that is linearly dependent on the concentration:
$$ D(\psi) = D_0 + D_1\psi$$

This choice represents the simplest non-trivial extension of constant diffusivity and has been observed in various experimental systems \cite{krishna1997maxwell, vazquez2006porous}. The parameter $D_1$ can be positive (enhanced diffusion at high concentrations) or negative (hindered diffusion), depending on the nature of particle interactions.

The resulting diffusion equation is:
$$\dfrac{\partial \psi}{\partial t} = D_0\dfrac{\partial^2 \psi}{\partial x^2} + \dfrac{D_1}{2}\dfrac{\partial^2 \psi^2}{\partial x^2}.$$

Here, $D_0$ is the standard Fickian diffusion coefficient, while $D_1$ accounts for the first-order correction due to concentration dependence.

The model is subject to an initial condition corresponding to a point source:
$$\psi(x,0)=m\delta(x),$$
where $m$ represents the total mass of the solute and $\delta(x)$ is the Dirac delta function. This initial condition is particularly relevant for understanding the fundamental spreading behavior from a localized source.

\subsection{Dimensionless Formulation}
To simplify the analysis, we introduce the following dimensionless variables:
$$\varphi(u,\tau)=\frac{1}{m}\psi(x,t),$$
$$u = \dfrac{x}{l},$$
$$\tau = \dfrac{D_0t}{l^2},$$
$$\varepsilon = \dfrac{mD_1}{2lD_0},$$
where $l$ is an arbitrary length scale. The dimensionless parameter $\varepsilon$ represents the strength of the nonlinearity relative to the linear diffusion term. In terms of these dimensionless variables, the nonlinear diffusion equation becomes:
$$\dfrac{\partial \varphi}{\partial \tau} = \dfrac{\partial^2 \varphi}{\partial u^2} + \varepsilon\dfrac{\partial^2 \varphi^2}{\partial u^2}.$$

This dimensionless formulation reveals that the behavior of the system is governed by a single parameter $\varepsilon$, which determines the relative importance of nonlinear effects.

% ==================================================
% Perturbative Analysis
% ==================================================
\section{Perturbative Analysis}
We employ a perturbation method to solve this equation, following the approach commonly used in nonlinear partial differential equations \cite{nayfeh1981introduction}. We assume the solution can be expanded as a power series in the perturbation parameter $\varepsilon$:
$$\varphi(u,\tau) = \sum_{n=0}^\infty\varepsilon^n\varphi_n(u,\tau).$$

Substituting this expansion into the dimensionless diffusion equation and collecting terms of the same order in $\varepsilon$ yields a hierarchy of linear partial differential equations:
\begin{itemize}
    \item Zeroth order ($\varepsilon^0$):
$$\dfrac{\partial \varphi_0}{\partial \tau} - \dfrac{\partial^2 \varphi_0}{\partial u^2} = 0$$

\item First order ($\varepsilon^1$):
$$\dfrac{\partial \varphi_1}{\partial \tau} - \dfrac{\partial^2 \varphi_1}{\partial u^2} = \dfrac{\partial^2 \varphi_0^2}{\partial u^2}$$

\item Second order ($\varepsilon^2$):
$$\dfrac{\partial \varphi_2}{\partial \tau} - \dfrac{\partial^2 \varphi_2}{\partial u^2} = 2\dfrac{\partial^2 (\varphi_0 \varphi_1)}{\partial u^2}$$
\end{itemize}

\subsection{Boundary and Initial Conditions}
The solutions $\varphi_n(u,\tau)$ must satisfy the following properties:
\begin{enumerate}
    \item Initial Condition: For $n>0$, $\varphi_n(u,0) = 0$. This ensures that the initial condition is fully satisfied by the zeroth-order solution.
    \item Mass Conservation: For $n>0$, $\int_{-\infty}^{\infty}\varphi_n(u,\tau)du = 0$. This ensures that mass is conserved at each order of the perturbation.
    \item Boundary Conditions: $\varphi_n(u,\tau) \to 0$ as $|u| \to \infty$ for all $n$ and $\tau > 0$.
\end{enumerate}

In general, for $n>0$, the equation for the $n$-th order correction is:
$$\dfrac{\partial \varphi_n}{\partial \tau} - \dfrac{\partial^2 \varphi_n}{\partial u^2} = F_n(u,\tau),$$

where the source term $F_n(u,\tau)$ is a function of the lower-order solutions and is defined as:
\begin{equation*}
    F_n(u, \tau) = \frac{\partial^2}{\partial u^2} \sum_{k=0}^{n-1} \varphi_k \varphi_{n-1-k}
\end{equation*}

This approach transforms the original nonlinear equation into an infinite set of coupled linear equations, each solvable using standard techniques for the heat equation.

\subsection{Solutions}
The solution for the zeroth-order equation ($n=0$) is the fundamental solution of the linear diffusion equation, also known as the heat kernel:
$$\varphi_0(u,\tau) = \dfrac{1}{\sqrt{4\pi\tau}}e^{-\dfrac{u^2}{4\tau}}.$$

The solutions for the higher-order equations ($n>0$) can be obtained using the Green's function method. The Green's function for the linear diffusion operator is:
$$G(u,\tau;z,\eta)=\varphi_0(u-z,\tau-\eta),$$

The solution for each $\varphi_n(u,\tau)$ is then given by the convolution:
$$\varphi_n(u,\tau) = \int_0^\tau \int_{-\infty}^\infty G(u,\tau;z,\eta) F_n(z,\eta)dzd\eta$$

Applying this method to find the first-order correction, $\varphi_1(u,\tau)$, yields:
$$\varphi_1(u,\tau) = -\dfrac{1}{4\pi\tau}e^{-\dfrac{u^2}{2\tau}} + \dfrac{u}{8\tau\sqrt{\pi\tau}}\text{erf}\left(\dfrac{u}{2\sqrt{\tau}}\right)e^{-\dfrac{u^2}{4\tau}}$$
where $\text{erf}(z)$ is the error function of $z$.

% ==================================================
% Results and Discussion
% ==================================================
\section{Results and Discussion}

\subsection{Diffusion Profiles}
\begin{figure}[H]
\centering
\includegraphics[width=\linewidth]{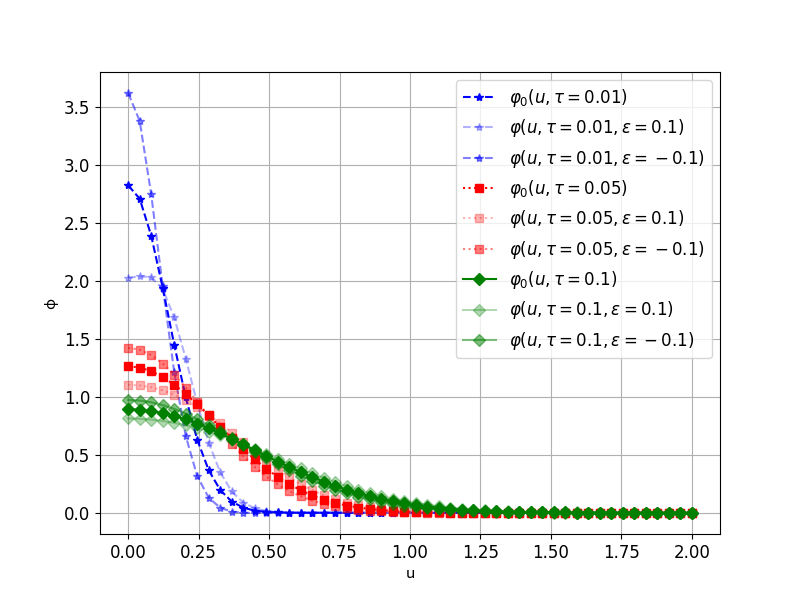}
\caption{Evolution of diffusion profiles for different values of the nonlinearity parameter $\varepsilon$. The profiles show how the concentration-dependent diffusivity affects the shape of the spreading distribution, with notable deviations from the Gaussian profile of linear diffusion.}
\label{fig:densities}
\end{figure}

The concentration profiles obtained from our perturbative analysis reveal interesting deviations from the standard Gaussian profile of linear diffusion. Figure \ref{fig:densities} shows the evolution of the diffusion profile for different values of the nonlinearity parameter $\varepsilon$.

The profiles demonstrate that the nonlinearity introduces asymmetries and changes in the peak height and width compared to the linear case. For positive $\varepsilon$ (enhanced diffusion at high concentrations), the profiles exhibit faster spreading, while negative $\varepsilon$ leads to hindered diffusion and narrower profiles.

\subsection{Mean Squared Displacement Analysis}
The mean squared displacement (MSD) is calculated as:
\begin{align}
    \langle u^2 \rangle &= \int_{-\infty}^\infty u^2 \varphi(u,\tau)du\\
                        &\approx  \int_{-\infty}^\infty u^2 \left(\varphi_0(u,\tau) + \varepsilon \varphi_1(u,\tau) \right)du
\end{align}

This yields:
$$\langle u^2 \rangle \approx 2\tau + \varepsilon\sqrt{\dfrac{2\tau}{\pi}}$$

The resulting MSD exhibits notable behavior comprising a term corresponding to classical Fickian diffusion ($2\tau$) and a first-order correction term that introduces anomalous behavior.

\subsection{Transient Anomalous Behavior}

\begin{figure}[H]
\centering
\includegraphics[width=\linewidth]{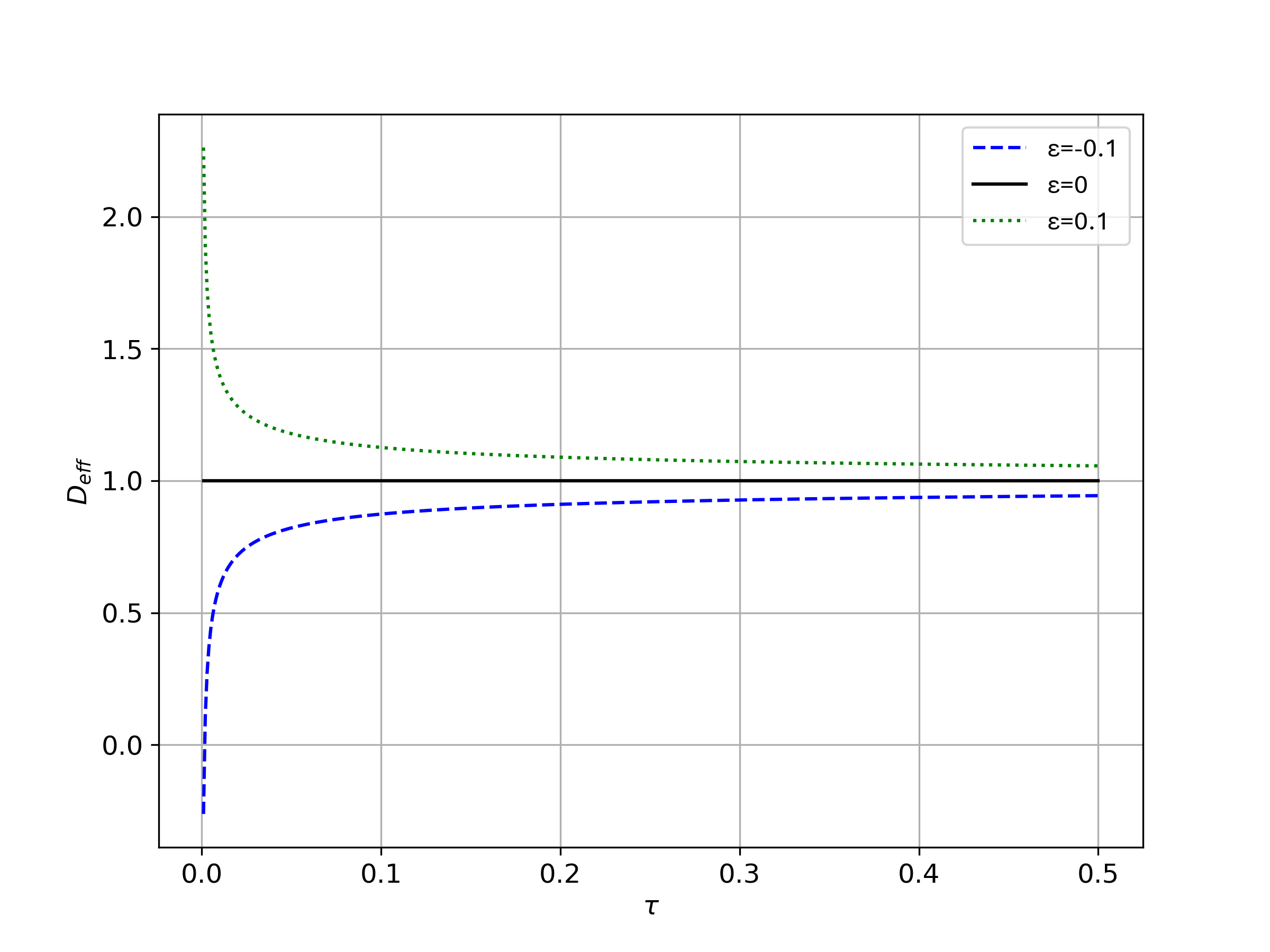}
\caption{Effective Diffusion Coefficient Demonstrating Transient Anomalous Behavior. This plot shows the effective diffusion coefficient $D_\text{eff}$ versus dimensionless time $(\tau)$for different values of the nonlinearity parameter, $\varepsilon$. The graph clearly illustrates the system's crossover from an initial anomalous diffusion regime to standard Fickian diffusion at long times.}
\label{fig:diffusion_coeff}
\end{figure}

The effective diffusion coefficient provides clear insight into the transient nature of the anomalous diffusion:
$$D_\text{eff}(\tau) = \frac{1}{2}\frac{d\langle u^2 \rangle}{d\tau} \approx 1 +\dfrac{\varepsilon}{2\sqrt{2\pi\tau}}$$
Figure \ref{fig:diffusion_coeff} illustrates this transient behavior by showing the normalized MSD as a function of time.

This figure demonstrates several key features:

\begin{itemize}
    \item Short-time behavior: At small $\tau$, the effective diffusion coefficient is dominated by the $\tau^{-1/2}$ term, leading to anomalous behavior with $\langle u^2 \rangle \propto \tau^{1/2}$.
    \item Long-time behavior: As $\tau$ increases, the linear term dominates, and the system recovers normal Fickian diffusion with $\langle u^2 \rangle \propto \tau$.
\end{itemize}

\subsection{Physical Interpretation}
The source term for the first-order correction, $F_1(u,\tau) = \frac{\partial^2 \varphi_0^2}{\partial u^2}$, has a zero spatial integral, indicating that the nonlinearity redistributes mass rather than creating or destroying it. Near the origin ($u=0$), $F_1$ acts as a sink, while at larger distances it acts as a source. This redistribution effect is responsible for the altered diffusion dynamics.

The physical mechanism can be understood as follows: in regions of high concentration (near the initial point source), the concentration-dependent diffusivity either enhances or hinders local diffusion depending on the sign of $D_1$. This creates a feedback effect where the local diffusion rate depends on the local concentration, leading to non-Gaussian spreading and transient anomalous behavior~\cite{cohen1959molecular,frank2005nonlinear}.

\subsection{Comparison with Other Models}
Our model provides several advantages over traditional approaches to anomalous diffusion:

\begin{enumerate}
    \item Physical transparency: Unlike fractional diffusion models~\cite{schneider1989fractional, meerschaert2008tempered}, our approach maintains clear physical interpretation through local interactions.
    \item Mathematical tractability: The perturbative approach allows for analytical insights into the mechanism of anomalous behavior.
    \item Experimental relevance: Concentration-dependent diffusivity has been observed in various experimental systems~\cite{banks2005Anomalous, jeon2016protein, sokolov2012models}.
\end{enumerate} 

However, it's important to note that our model predicts only transient anomalous behavior, unlike some systems that exhibit persistent anomalous diffusion over extended time ranges~\cite{jeon2011in}.

\subsection{Limitations and Validity}
The solution presented is a perturbative expansion, formally accurate for small values of the nonlinearity parameter ($\epsilon \ll 1$). However, to fully delineate the physical validity of the model, it is instructive to analyze the dimensional mean squared displacement:
$$\langle x^2(t) \rangle \approx 2 D_0 t + m D_1  \sqrt{\frac{t}{\pi D_0}}.$$
This expression reveals that the anomalous correction scales with the total mass $m$, implying that the approximation is most robust for systems where collective effects are moderate. The sign of the interaction parameter $D_1$ clearly distinguishes two physical regimes: hindered diffusion ($D_1 < 0$), characteristic of crowded environments or dense liquids (e.g., Lennard-Jones fluids) where "caging" effects dominate; and enhanced diffusion ($D_1 > 0$), typical of systems driven by repulsive forces or pressure gradients, such as gas expansion.

A critical limitation of the first-order truncation arises in the short-time limit ($t \to 0$). For hindered diffusion ($\epsilon < 0$), the anomalous term ($\propto \sqrt{t}$) mathematically dominates the linear Fickian term ($\propto t$), potentially yielding unphysical results (such as negative MSD values) due to the breakdown of the perturbative approximation at $t \to 0$. This behavior is an artifact of the perturbative approximation near the initial singularity, where local concentration gradients are infinite. Consequently, our analytical results are strictly valid for intermediate times—after the initial singularity has relaxed but before the system becomes fully dilute.

Finally, our analysis is currently restricted to one dimension and specific initial conditions. Extensions to higher dimensions or systems with strong nonlinearities would likely require non-perturbative numerical simulations to fully capture the dynamics \cite{chukbar1995stochastic}.

\section{Conclusion}
In this work, we have demonstrated that a simple, physically-grounded nonlinear diffusion equation can give rise to transient anomalous diffusion. By considering a diffusion coefficient with a linear dependence on concentration, our perturbative analysis shows a clear crossover in the diffusive behavior. The system initially exhibits subdiffusion, with a mean squared displacement proportional to $\sqrt{t}$, before transitioning to classical Fickian diffusion, where the MSD is proportional to $t$, in the long-time limit.

The significance of this result lies in its mechanism. The model provides a compelling alternative to tempered anomalous diffusion for explaining transient anomalies in certain systems. Instead of relying on non-local interactions or memory effects, which can be challenging to justify physically, our model attributes the anomalous behavior to purely local, concentration-dependent interactions. This suggests that such transient phenomena may not require exotic underlying physics but can emerge naturally from the collective behavior in crowded or interacting particle systems.

\end{multicols}

\bibliographystyle{unsrt}
\bibliography{ref}

\end{document}